\documentclass[finalold]{agujournal2019}
\usepackage{url} 
\usepackage{lineno}
\usepackage[finalnew]{trackchanges} 
\usepackage{soul}

\usepackage[hidelinks,backref]{hyperref}
\hypersetup{linktoc=all,colorlinks=true,linkcolor={blue!80!black},citecolor={blue!80!black}}
\usepackage{orcidlink}

\usepackage{graphicx}	
\usepackage{amsmath}	
\usepackage{textcomp,gensymb}	
\usepackage{overpic}

\newcommand{\isois}{IS\raisebox{1.1 pt}{$\odot$}IS}
\newcommand{\app}{$\sim$}
\newcommand{\gt}{\textgreater}
\newcommand{\lt}{\textless}
\newcommand{\kms}{$km s^{-1}$}

\addeditor{rev2}

\newcommand{\grn}{\textcolor{teal}}

\DeclareUnicodeCharacter{2212}{\ensuremath{-}}
\DeclareUnicodeCharacter{2264}{\ensuremath{-}}

\usepackage{natbib}

\draftfalse

\journalname{Space Weather}

\begin{document}
\begin{sloppypar}

%
%


\title{\change{Multi-spacecraft observations of e}{E}nergy-dependent SEP Fe/O abundances during the May 2024 superstorm}
%
%




\authors{G.D. Muro\affil{1}\orcidlink{0000-0003-0581-1278}, 
C.M.S. Cohen\affil{1}\orcidlink{0000-0002-0978-8127}, 
Z. Xu\affil{1}\orcidlink{0000-0002-9246-996X}, 
R.A. Leske\affil{1}\orcidlink{0000-0002-0156-2414}, 
A.C. Cummings\affil{1}\orcidlink{0000-0002-3840-7696}, 
S. Bale\affil{2}\orcidlink{0000-0002-1989-3596}, 
G. D. Berland\affil{3}\orcidlink{0000-0001-6010-6374}, 
E. R. Christian\affil{4}\orcidlink{0000-0003-2134-3937}, 
M. E. Cuesta\affil{5}\orcidlink{0000-0002-7341-2992}, 
M. I. Desai\affil{6,7}\orcidlink{0000-0002-7318-6008}, 
F. Fraschetti\affil{8,9}\orcidlink{0000-0002-5456-4771}, 
J. Giacalone\affil{9}\orcidlink{0000-0002-0850-4233}, 
L. Y. Khoo\affil{5}\orcidlink{0000-0003-0412-1064}, 
A. Labrador\affil{1}\orcidlink{0000-0001-9178-5349}, 
D. J. McComas\affil{5}\orcidlink{0000-0001-6160-1158}, 
J. G. Mitchell\affil{4}\orcidlink{0000-0003-4501-5452}, 
M. Pulupa\affil{2}\orcidlink{0000-0002-1573-7457}, 
N. A. Schwadron\affil{10}\orcidlink{0000-0002-3737-9283}, 
M. M. Shen\affil{5}\orcidlink{0000-0002-3093-458X}} 

\affiliation{1}{California Institute of Technology, Pasadena, CA 91125, USA}
\affiliation{2}{University of California Berkeley, Berkeley, CA 94720, USA}
\affiliation{3}{Johns Hopkins University Applied Physics Laboratory, Laurel, MD 20723, USA}
\affiliation{4}{NASA/Goddard Space Flight Center, Greenbelt, MD 20771, USA}
\affiliation{5}{Department of Astrophysical Sciences, Princeton University, Princeton, NJ 08544, USA}
\affiliation{6}{Southwest Research Institute, San Antonio, TX 78228, USA}
\affiliation{7}{University of Texas, San Antonio, TX 78249, USA}
\affiliation{8}{Center for Astrophysics $|$ Harvard $\&$ Smithsonian, Cambridge, MA 02138, USA}
\affiliation{9}{University of Arizona, Tucson, AZ 85721, USA}
\affiliation{10}{University of New Hampshire, Durham, NH 03824, USA}







\correspondingauthor{Gabriel D. Muro}{gmuro@caltech.edu}






\begin{keypoints}
\item SEP composition analyzed across 90\degree\ longitude using PSP, STEREO-A, and ACE during May 2024 storm from AR 13664.

\item All spacecraft observed Fe/O ratios rising with energy, inconsistent with shock acceleration predictions.

\item Stepwise SEP rises and delayed onsets suggest CME mergers and SEP reacceleration shaped ion fluence and composition.
\end{keypoints}

%
%

%
%


\begin{abstract}

During mid-May 2024, active region (AR) 13664 produced a series of M- and X-class flares along with several coronal mass ejections (CMEs) that resulted in exceptionally strong aurora at Earth. This study presents in-situ solar energetic particle (SEP) ion composition data from Solar Terrestrial Relations Observatory Ahead (STA), Advanced Composition Explorer (ACE), and Parker Solar Probe (PSP) as their magnetic connectivity to AR 13664 varied throughout the event period. Between 08 to 24 May, STA was separated by 12\degree\ in longitude from ACE at 0.96 AU. SEP intensities rose gradually due to merged CMEs from AR 13664. On 13 May, an M6 flare was followed by a rapid-onset SEP event at STA, although velocity dispersion analysis yielded no clear path length or release time. PSP, ~95\degree\ longitudinally separated from Earth at 0.74 AU, observed gradually increasing SEP intensities beginning 11 May, followed by a jump in both SEP intensity and magnetic field (\gt100 nT) on 16 May. These early event intervals display stepwise SEP increases, consistent with the passage of successive CMEs. On 20 May, an X16.5 flare from AR 13664 produced an Fe-rich SEP event observed at all three spacecraft despite their wide longitudinal separations. Throughout the period, Fe/O ratios ranged from \lt0.01 to \gt0.8 and increased with energy between 1 to 100 MeV/nuc. This trend deviates from the typical energy-dependent decrease expected from diffusive shock acceleration and suggests more complex scenarios, possibly involving variable suprathermal seed populations or species-dependent transport.

\end{abstract}


\section*{Plain Language Summary}

In May 2024, the Sun produced a series of powerful solar flares and coronal mass ejections (CMEs) from AR 13664. These events led to spectacular aurorae on Earth and launched solar energetic particles (SEPs) into space. We used data from three spacecraft, STEREO-A, ACE, and Parker Solar Probe, that were positioned in different locations around the Sun to study how these particles spread through the solar system. 

The spacecraft observed increases in SEP intensity at different times, depending on how well they were magnetically connected to AR 13664. In some cases, the particle increases happened in steps, likely caused by multiple CMEs combining in space. On \change{May 20}{20 May}, a particularly powerful flare sent an iron-rich SEP event in all directions, reaching all spacecraft\add{, despite} their wide separation.

We also measured how the mix of ions in SEPs, especially the ratio of iron to oxygen, changed with energy. Normally, we expect this ratio to decrease with energy, but during this event it increased. This unusual behavior suggests that the standard models of particle acceleration may not fully explain what occurred. Instead, more complex factors, such as the relative ionization of the SEPs or how \change{they gain energy during transport}{their energy evolves due to transport effects}, are discussed.

%
%

%


%
%
%
%

\section{Introduction}\label{introduction}

Solar superstorms, characterized by phenomena such as strong solar flares, rapid sequences of coronal mass ejections (CMEs), and intense solar energetic particle (SEP) events represent some of the most dynamic processes in our solar system. These \change{superstorms}{so-called "superstorms"} showcase the Sun's \add{most} powerful behavior and also pose significant risks to Earth's technological infrastructure and space operations \citep{eastwood2018}, \add{particularly when} their Disturbance Storm Time index \change{reaches}{falls to} $\leq -250$ nT \citep{meng2019}\add{, marking dramatic impacts on the geospace envionment}. Historically, superstorms like the 1859 Carrington Event, 2000 Bastille Day Event, 2003 Halloween Events, and 2015 St Patricks Day Events have exhibited the potential for solar activity to disrupt life on Earth \citep{carrington1859,watari2001,gopalswamy2005,wu2016}, emphasizing the urgent need for understanding and predicting such occurrences.

The period of mid-May 2024 had intense solar activity centered around Active Region (AR) 13664. This AR was responsible for producing a series of M- and X-class flares, along with associated CMEs, which led to significant geomagnetic storms and aurorae at Earth. After AR 13664 rotated to the far side of the Sun, on 20 May 2024 it produced an X16.5 flare (as estimated by Solar Orbiter/STIX) \citep{stiefel2025}, the most powerful observed since the launch of Parker Solar Probe (PSP) \citep{fox2016} in 2018 and resulted in the largest SEP event ever recorded on the surface of Mars \citep{lowe2025}.

During superstorms, it is challenging to distinguish one SEP accelerating mechanism from another due to an overwhelming amount of solar activity. The ion composition of SEP events can provide some clues regarding the environment in which acceleration processes occur. For instance, high iron-to-oxygen (Fe/O) ratios often indicate that particles have been accelerated to MeV/nuc energies in regions close to the Sun, such as near a flare site \citep{li2005a}. Conversely, low Fe/O ratios suggest interplanetary shocks driven by CMEs or a longer transport path through interplanetary space \citep{li2005a,cane2006,park2024}.

\add{SEPs are accelerated primarily by flare processes in the low corona and by diffusive shock acceleration (DSA) at CME-driven shocks} \citep{cane2006,desai2016}\add{. In DSA, fluence spectra are often described by power laws with a spectral break above a characteristic energy where acceleration becomes inefficient. Models predict ion species to share similar spectral slopes below this break, yielding Fe/O approximately energy independent. The break energy is expected to be element dependent, with lower energy values for higher mass elements, and thus the Fe/O abundance would decrease above the break. Flare-related seed populations, by contrast, often show enhanced heavy-ion abundances and higher charge states, with "impulsive" events Fe/O ratio rising above 0.5, compared to the average solar wind Fe/O ratio of 0.134 observed in large shock-accelerated events at 5–12~MeV/nuc} \citep{reames1995,mason2004}\add{. Event-to-event variability at higher energies can reflect the interplay of shock geometry and seed populations }\citep{tylka2006}. 

\add{Transport processes can further influence SEP observations, as pitch-angle scattering and field-line meandering can delay onsets and alter velocity-dispersion signatures, while cross-field diffusion and drifts introduce rigidity- and species-dependent propagation effects} \citep{giacalone1999,dalla2013}\add{. \change[rev2]{In the heliosphere, the solar wind and magnetic focusing (adiabatic deceleration) shift spectra to lower energies, while transient structures such as CMEs modify magnetic connectivity, imprinting additional rigidity- and species-dependent delays}{In a simple Parker-spiral magnetic field and a radial constant solar wind, any charged particle will lose energy with time }\citep{kota2000}\add[rev2]{ in a process referred to as adiabatic deceleration. This also occurs when pitch-angle scattering and spatial diffusion in turbulent magnetic fields occurs, and is an important transport process in the heliosphere }\citep{jokipii1971}\add[rev2]{. In addition, transient structures, such as CMEs, can modify magnetic connectivity to acceleration sources, imprinting rigidity- and species-dependent delays in arrival time}} \citep{desai2016}\add{. Early in events, large intensity gradients make perpendicular diffusion particularly effective, broadening longitudinal spread} \citep{strauss2017}\add{. Finally acceleration at shocks is ultimately limited due to particles escaping from the shock, which is generally related to Q/M-dependent scattering. This produces breaks in the power law spectra that may reflect shock geometry or propagation speed but also the Q/M dependence} \citep{schwadron2015,fraschetti2021}\add{. Thus, interpretations of the Fe/O energy dependence require accounting for both source acceleration and transport effects.}

\add{Previous composition studies have shown strong variability with longitude and connectivity. Multi-spacecraft observations with Advanced Composition Explorer (ACE)} \citep{stone1998} \add{and Solar Terrestrial Relations Observatory Ahead (STA)} \citep{kaiser2007} \add{demonstrated that heavy-ion fluence distributions can span broad longitudinal ranges, in ways not always explained by transport alone} \citep{lario2013,cohen2021,kouloumvakos2024,muro2025}\add{. This study extends this approach by combining PSP, STEREO-A, and ACE, covering over} \app90\degree \add{ in longitude during the May 2024 superstorm period. We show that Fe/O seems to increase with energy across multiple vantage points, a departure from} \remove{idealized }\add{DSA expectations, with implications for the role of flare-related seed populations, quasi-perpendicular shocks, and transport effects.}

\remove{The May 2024 period provided a unique opportunity to perform multi-spacecraft analysis with PSP, Solar Terrestrial Relations Observatory Ahead (STA), and Advanced Composition Explorer (ACE) to examine the longitudinal variance of Fe/O and complexities in magnetic connectivity and particle transport. This study explores the ion composition of three SEP event periods using the wide longitudinal separation of the spacecraft during this period of strong geomagnetic activity,}

\section{Spacecraft configuration and solar activity}\label{spacecraft}

\begin{figure}[htbp]
    \centering
    \includegraphics[width=0.99\linewidth]{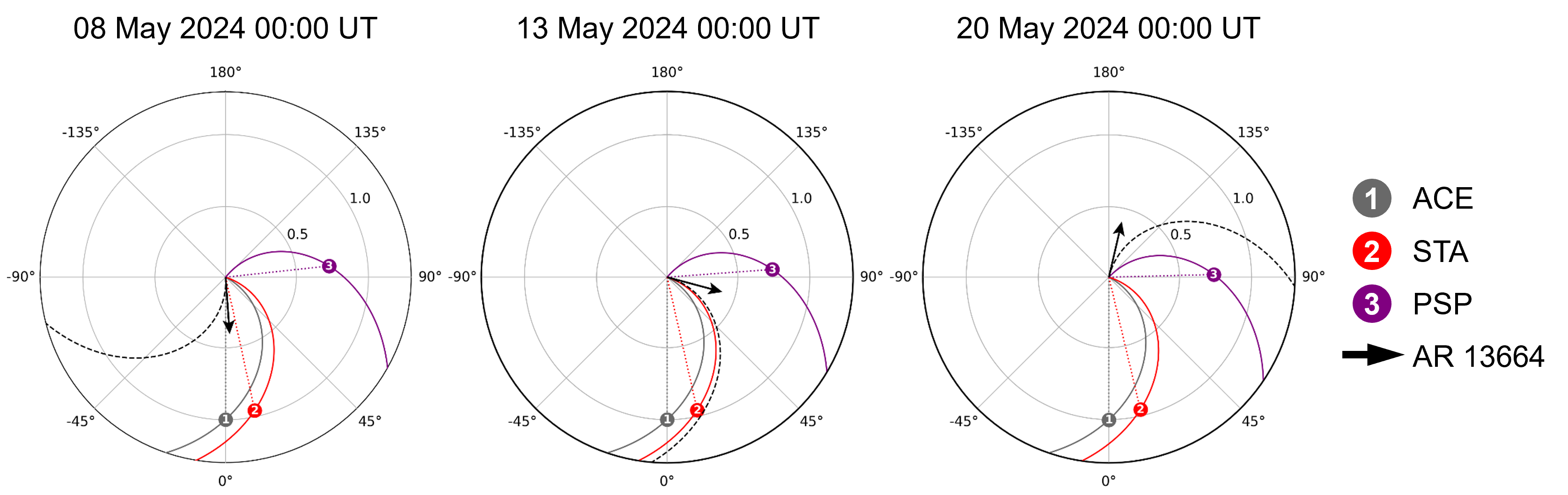}
    \caption{Spacecraft configuration of ACE in grey, STA in red, and PSP in violet from the Solar-MACH model \citep{gieseler2023} during mid-May 2024. The black reference arrows designate the longitude of AR 13664. All nominal Parker spiral field lines are \change{projected to 1 AU}{mapped back from the observer} at (left) 00:00 on 8 May, (middle) 00:00 on 13 May and (right) 00:00 on 20 May, based on 400 \kms solar wind speed, and colored to match their associated spacecraft, with the dotted line corresponding to the AR field line. See Table \ref{tab-solarmach} for \change{specifications}{coordinates}.}
    \label{fig-solarmach}
\end{figure}

 \begin{table}
 \caption{Stonyhurst coordinates of spacecraft in Figure \ref{fig-solarmach} for 8 \remove{May 2024 }to 24 May 2024\add{ and their sub-solar longitudinal separation from AR 13664 at each time shown in Figure {\ref{fig-solarmach}}.}}
 \centering
 \begin{tabular}{l c c c}
 \hline
  & PSP & STA & ACE \\
 \hline
   Stonyhurst longitude [\degree] & 96.2 to 90.3 & 12.3 to 13.6 & 0.0 to -0.1 \\
   Stonyhurst latitude [\degree] & 3.6 to 3.8 & -1.9 to 0.1 & -3.4 to -1.7 \\
   Heliocentric distance [AU] & 0.73$^{a}$ & 0.96 & 1.0 \\
   8 May 2024 separation [\degree] & 85.5 & 7.4 & -6.0 \\
   13 May 2024 separation [\degree] & 16.5 & -61.6 & -75.0 \\
   20 May 2024 separation [\degree] & -77.5 & -155.6 & -169.0 \\
 \hline
 \multicolumn{2}{l}{$^{a}$Aphelion of 0.7448 AU on \change{May 15}{15 May}.}
 \end{tabular}
    \label{tab-solarmach}
 \end{table}

Figure \ref{fig-solarmach} \add{and Table {\ref{tab-solarmach}} }displays the relative locations of PSP, STA, and ACE at the \remove{beginning of each interval} from 8 to 20 May 2024 at 0:00 UT, these dates were chosen due to the SEP events associated with AR 13664. During this study period, \add{coordinates in Table {\ref{tab-solarmach}} show that }all three spacecraft stayed at nearly the same radial distance from the Sun and their longitudinal separation varied only slightly. Thus, the primary change in relative position with respect to AR 13664, located at -20\degree\ latitude and ranging in longitude from -4\degree\ to -162\degree\ in Stonyhurst coordinates \citep{thompson2006}, was due to solar rotation.

\begin{figure}[htbp]
    \centering
    \includegraphics[width=0.7\linewidth]{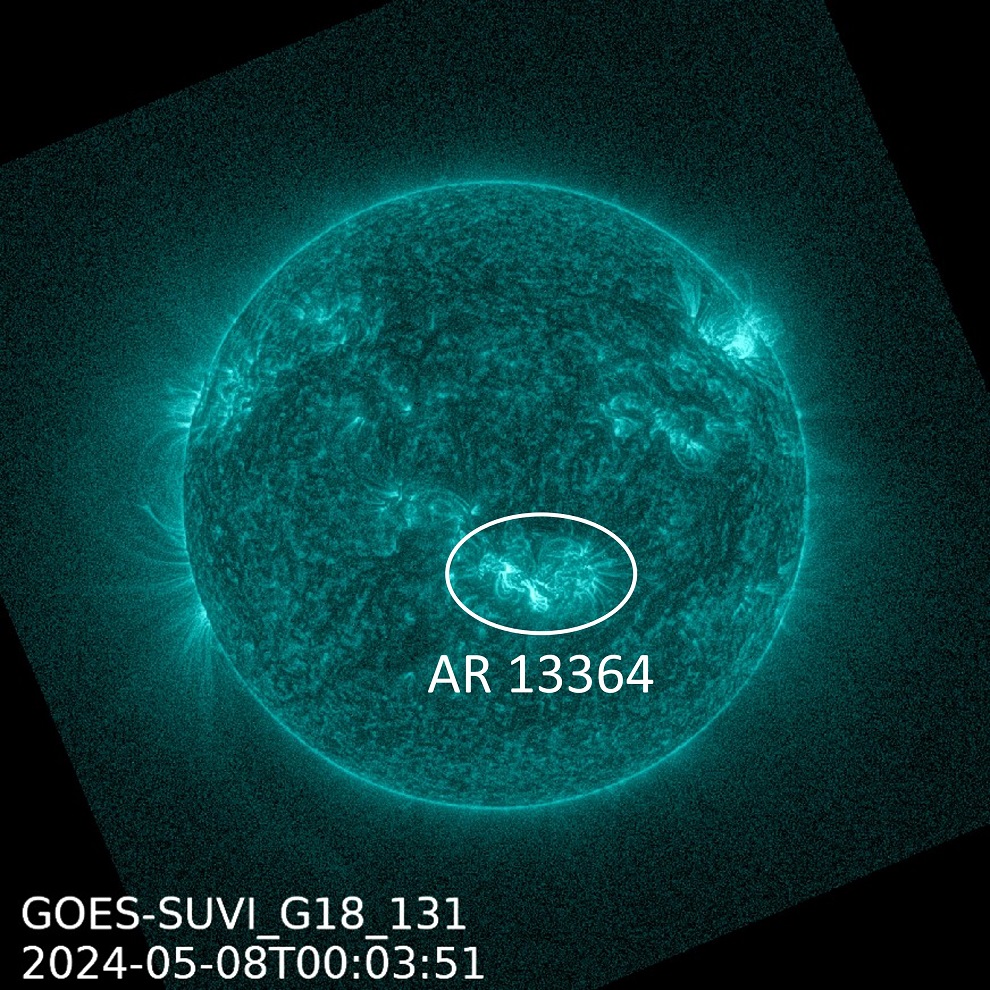}
    \caption{Ultraviolet image from GOES/SUVI 131 on 8 May 2024, enhanced with multi-scale Gaussian normalization.}
    \label{fig-hcs_suvi}
\end{figure}

AR 13664 formed on the far side of the Sun in April 2024 and was visible from Earth from \app1 to 13 May 2024. Figure \ref{fig-hcs_suvi} shows the location of AR 13664 on 8 May at 00:03 UT from the GOES-18 spacecraft located near Earth. The strength and abundance of solar flares emitted from AR 13664 was especially notable, with 12 X-class flares measured by GOES/SXR from 8 May until 15 May when the AR had fully rotated over the limb from Earth's field of view \citep{li2024}. On the solar farside, AR 13664 continued to produce 5 more X-class flares through 26 May and returned to the nearside with 6 more X-class flares through 11 June \citep{jaswal2025}. This AR was characterized by its extremely large size and total unsigned line-of-sight magnetic flux, ranking it among the top 0.1\%\ of historical ARs observed since 1874 in the MDI/HMI and RGA/NOAA databases. Its evolution included merging with AR 13668, which originated from the same underlying magnetic flux system \citep{jaswal2025}.

From 8 May until 15 May, there were 10 CMEs associated with the aforementioned X-class flares and another 10 associated with C- and M-class flares, at vastly different speeds from 311 to 2011 \kms. Three distinct clusters between 10-12 May, with eruptions as short as 14 minutes apart, arrived at Earth despite a significantly larger number observed via coronagraph \citep{hayakawa2025}. Therefore, it is understood that multiple CMEs merged on the way to Earth making it difficult to connect any specific CME observed in the coronagraph data to the energetic particles measured in situ. 

\section{In-situ SEP instruments}\label{insitu_instrument}

This study utilizes energetic particle instruments on PSP, STA, and ACE that measure a similar energy range to investigate SEPs during the May 2024 period. Each spacecraft employs distinct instrument suites tailored for varying energy ranges, mass resolutions, and spatial orientations. While for this study PSP provides observations at \app0.73 AU and 90 to 96\degree\ ahead of Earth, STA offers a longitudinal perspective at 1 AU, \app13\degree\ ahead of Earth, and ACE contributes measurements at the L1 position. This combination of in-situ instruments allows for monitoring of SEPs emitted from AR 13664 as it rotates beneath their positions and the connectivity between the region and each spacecraft changes. 

The Integrated Science Investigation of the Sun (\isois) onboard PSP is a state-of-the-art particle detection suite designed to study high energy particles in the inner heliosphere \citep{mccomas2016}. The suite consists of two instruments, of which only one is used in this study to match the energy range of its counterparts on the other spacecraft: EPI-Hi, a high-energy detector measuring \change{particles}{ions} in the range of \app1 MeV/nuc to $\geq$100 MeV/nuc. EPI-Hi is subdivided into the Low-Energy Telescope (PSP/LET), which covers energies from \app1 to 20 MeV/nuc and the High-Energy Telescope (PSP/HET), which measures \gt10 MeV/nuc ions. Both utilize the dE/dx versus residual energy technique \citep{wiedenbeck2017}, which measures how much energy a charged particle loses passing through each detector layer compared to how much energy is deposited into a subsequent detector layer. For this study, we present H, He, O, and Fe intensities from the sunward-facing aperture of PSP/LET, and H intensities from the sunward-facing aperture of PSP/HET, centered 45\degree\ and 20\degree\ off the spacecraft-Sun line, respectively.

The STA spacecraft carries the Low Energy Telescope (STA/LET) instrument to measure energetic particles with energies ranging from 1 to \gt20 MeV/nuc \citep{mewaldt2007}, depending on ion species, and the High-Energy Telescope (STA/HET) instrument to measure H ions from 3 to 30 MeV/nuc \citep{vonRosenvinge2008}. The ACE spacecraft is equipped with the Solar Isotope Spectrometer (ACE/SIS) \citep{stone1998}, which measures SEPs from \app10 to 100 MeV/nuc. All in-situ instruments in this study employ the dE/dx versus residual energy technique to provide ion composition data, albeit with differing energy bins. PSP/LET and STA/LET have measurements of H, He, O, and Fe, while ACE/SIS provides measurements of He, O, and Fe. The time cadence of PSP/HET, PSP/LET, STA/HET, and STA/LET is 60 seconds and ACE/SIS is slower with intensity measurements available every 256 seconds.

\section{Energetic particle measurements}\label{energetic_particle_measurements}

\subsection{Solar interval partitioning}\label{solar_storm_partition}

\begin{figure}[htbp]
    \centering
    \includegraphics[width=1.0\linewidth]{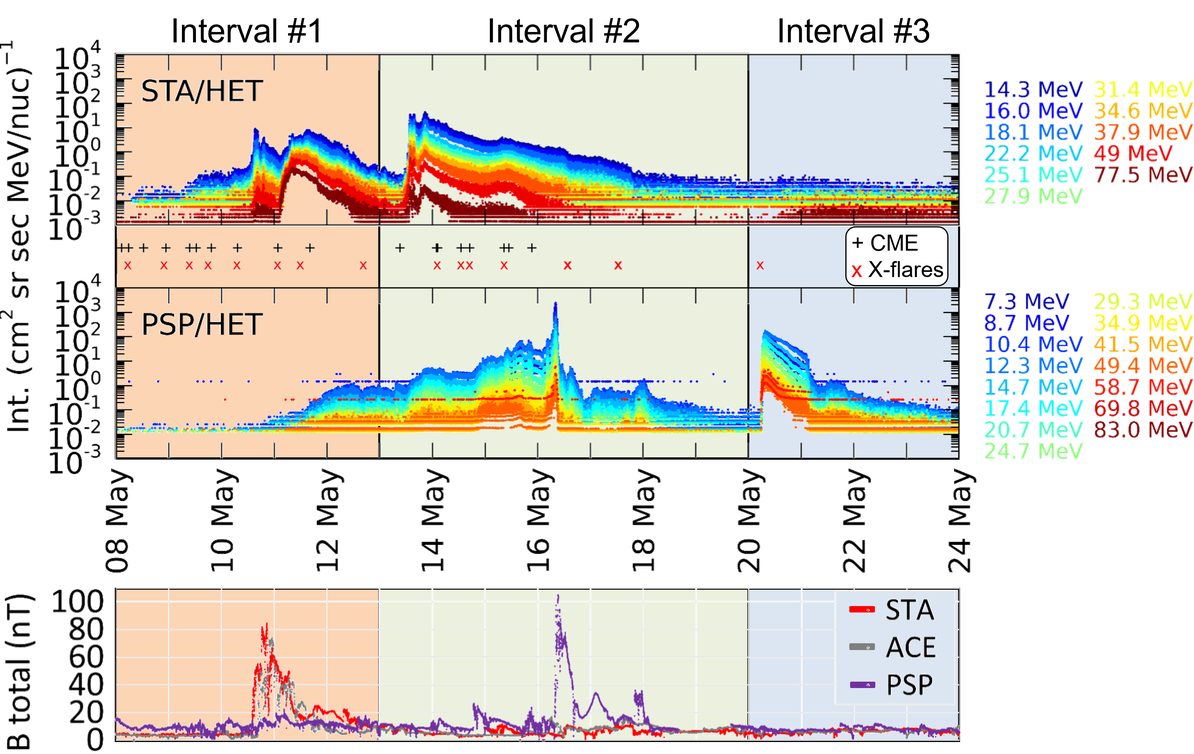}
    \caption{(Top) Proton measurements from STA/HET and PSP/HET, partitioned into three separate periods of solar intervals, from 8 to 24 May 2024. Between them is a visual representation of the timings for CMEs as a black cross and X-class flares as a red X generated by AR 13664, adapted from \citet{hayakawa2025} and \citet{jaswal2025}. (Bottom) Magnetic field measurements from ACE, STA, and PSP.}
    \label{fig-storms}
\end{figure}

Energetic protons were recorded from 8 through 24 May 2024. Based on the time profile of the SEP data, we decided to separate them into 3 different temporal periods, according to the intensity jumps measured at STA/HET and PSP/HET called Interval 1, 2, 3, as shown in Figure \ref{fig-storms}. Note that ACE/SIS does not measure protons, so only PSP/HET and STA/HET are shown. \add{Figure {\ref{fig-storms}} also displays the time of eruption for 17 X-class flares and 18 CMEs, which were adapted from lists in }\citet{hayakawa2025} and \citet{jaswal2025}\add{. Although, it is important to note that CMEs were no longer tracked after AR 13664 rotated away from the Earth.} Exceptionally strong CME passages occurred in the first two intervals, and a strong flare with a fast CME along with a rapid onset SEP event marks the third interval for PSP.

Figure \ref{fig-storms} shows the light orange-colored Interval 1, covering 8 through 12 May, with most activity occurring at STA. ACE displays similar activity in Figure \ref{fig:ace_sis} in heavier ions as well. The general SEP flux evolution at STA is marked by step increases in proton intensity. 

The high number of flares and eruption of 6 CMEs generally increased the overall SEP proton environment until 10 May, when \change{B$_{total}$ field}{the total magnetic field strength (B$_{total}$, the magnitude of the three-component magnetic field vector)} rapidly increased for STA and ACE around 17:00 UT. The 88 nT measurement at STA was among the strongest ever measured by the spacecraft \citep{russell2013}. Following the passage of these merged CMEs, the proton intensity sharply increased on 10 May and generally remained elevated for STA, located 12.5\degree\ ahead of Earth through 11 May. The \change{day long}{day-long} high B$_{total}$ from 10 to 11 May coincided with the geoeffective period that resulted in aurorae appearing at low latitudes on Earth and similarly sharp increases in He intensity for both STA and ACE, as shown in Figures \ref{fig:stereo_let} and \ref{fig:ace_sis}. PSP, oriented nearly 90\degree\ heliographic longitude ahead of the other two spacecraft, showed comparatively little B$_{total}$ activity with minor peaks during the two-day geoeffective period. Curiously, Figure \ref{fig:psp_let} shows that PSP/LET measured a slight decrease in protons during this time and gradually increased in proton intensity towards the end of 11 May. 
Figure \ref{fig-storms} shows the light green-colored Interval 2, covering 13 through 19 May. B$_{total}$ decreases below 20 nT for both STA and ACE, while the most distinctive feature for STA/HET is a rapid increase of proton flux on 13 May around 12:45 UT. Earlier that same day, an M6.6 flare began at 8:48 and peaked at 09:45 UT with elevated X-ray flux for \app8 hours as well as a halo CME \add{traveling at 1700 \kms} when AR 13664 was located at -23\degree, 80\degree\ \citep{liu2024}. The nearly 3 hour gap between flare eruption and rapid onset of STA/HET proton intensity suggests that the SEP increase was unlikely to be the result of this flare, whereas PSP showed distinctly different SEP profiles and magnetic field behavior. Notably, there are several jumps in B$_{total}$ from 3 to 27 nT which suggests the passage of CMEs at PSP's location, with small jumps in proton intensity occurring throughout the Interval 2 period prior to 16 May. The largest increase in B$_{total}$ to 108 nT over the entire study period occurs on 16 May and coincides with both dynamic threshold \add{(DT)} modes 1 and 2 being engaged \citep{cohen2021} for PSP/LET starting at 07:03 UT when the largest proton intensity during this study occurred for PSP. PSP/HET also engaged DT mode 1 at 07:26 UT during this same period. \add{DT modes are on-board, rate-dependent trigger modes that engage when the SEP count rate crosses a limit; the instrument raises energy thresholds on specific silicon detector segments to keep livetime high, reduce geometry factor, but do not affect the heavy ion intensities which are the key analysis in this study. In PSP/LET, DT mode 1 reduces the geometry factor and measurements below 2 MeV/nuc for protons and He, while in PSP/HET the geometry factor is reduced with no effect on the energy range coverage. In PSP/LET, DT mode 2 instrument response remains to be completely characterized, so the protons and He intensities are not included in Figure {\ref{fig-fluence}}.} Following this peak of SEP activity, another CME crossing occurred at 20:00 on 17 May with enhanced proton intensity. After that, a \change{day long}{day-long} data gap in B$_{total}$ for PSP ended the period. Proton intensity for 18 and 19 May showed a gradual decrease without any notable events occurring for PSP or STA.

Figure \ref{fig-storms} shows the light blue-colored Interval 3 period occurs when AR 13664 has already fully moved towards the farside of both STA and ACE, but was near the southwest limb from PSP's point of view with good nominal Parker spiral connectivity for a 400 \kms\ solar wind speed. On 20 May at 05:12 UT, an exceptionally strong Type III radio burst was measured by PSP with a Type II radio burst becoming distinguishable \app8 minutes later. The STIX instrument onboard Solar Orbiter observed an X16.5 flare occurring from AR 13664 \citep{stiefel2025}, which was the strongest flare measurement since PSP's launch in 2018\add{, and released a halo CME}. These radio and X-ray signatures were soon followed by an impulsive arrival of proton intensity at PSP/HET on 05:42 UT\change{.}{ and more gradual arrival of protons at 07:40 UT at STA/HET, shown in Figure {\ref{fig-storms}}. In Figure {\ref{fig:stereo_let}}, STA/LET also shows an increase in proton intensity as well as a slight amount in Fe later on 20 May. In Figure {\ref{fig:ace_sis}}, ACE/SIS also shows an increase in Fe later in the day, suggesting this SEP event was not longitudinally narrow.}

\subsection{Ion composition measurements}\label{ion_composition_measurements}

\begin{figure}[htbp]
    \centering
    \includegraphics[width=1.0\linewidth]{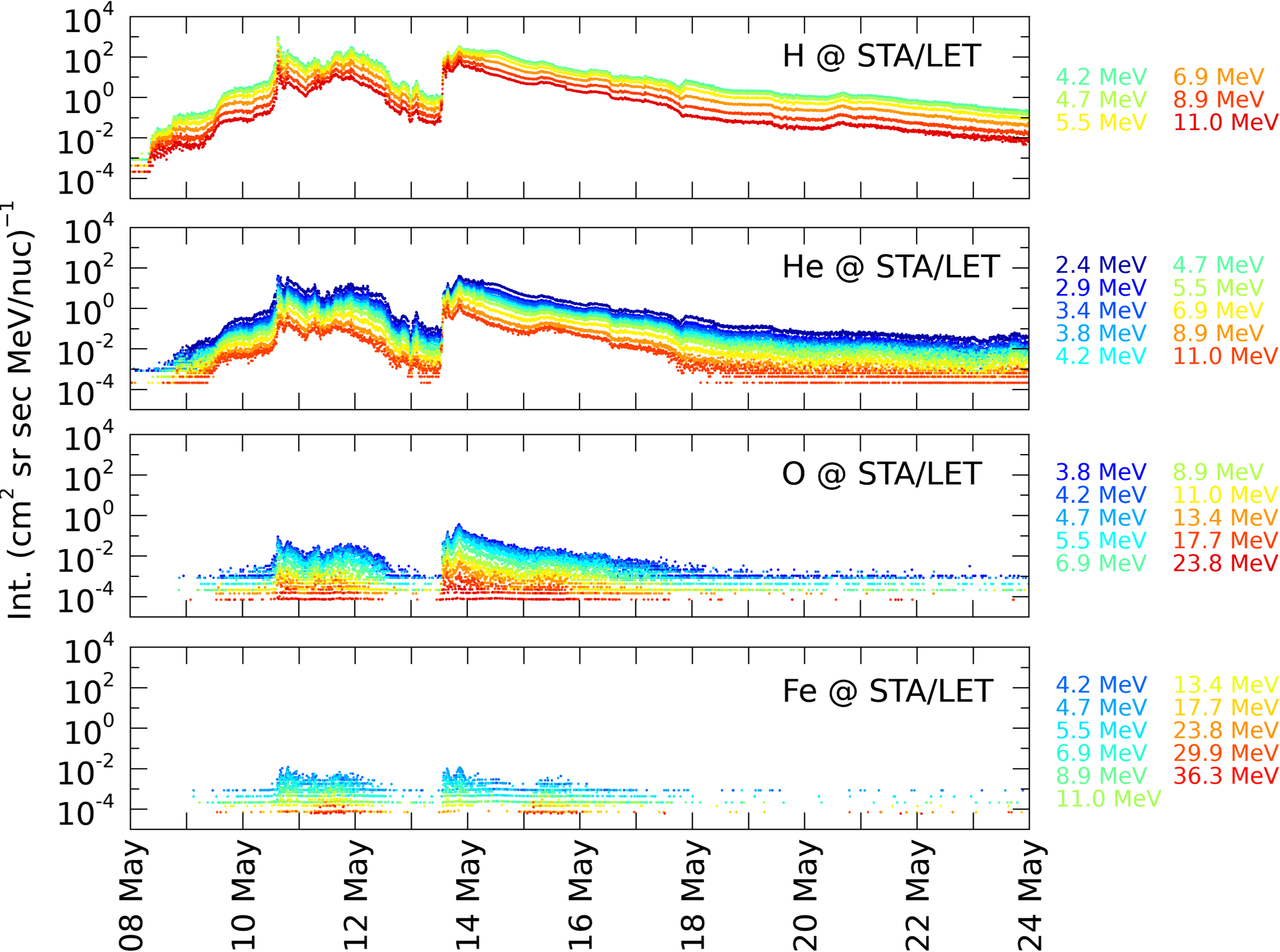}
    \caption{Intensity plots measured by STA/LET A for: (1) protons, (2) helium, (3) oxygen, and (4) iron at 1 minute cadence.}
    \label{fig:stereo_let}
\end{figure}

\begin{figure}[htbp]
    \centering
    \includegraphics[width=1.0\linewidth]{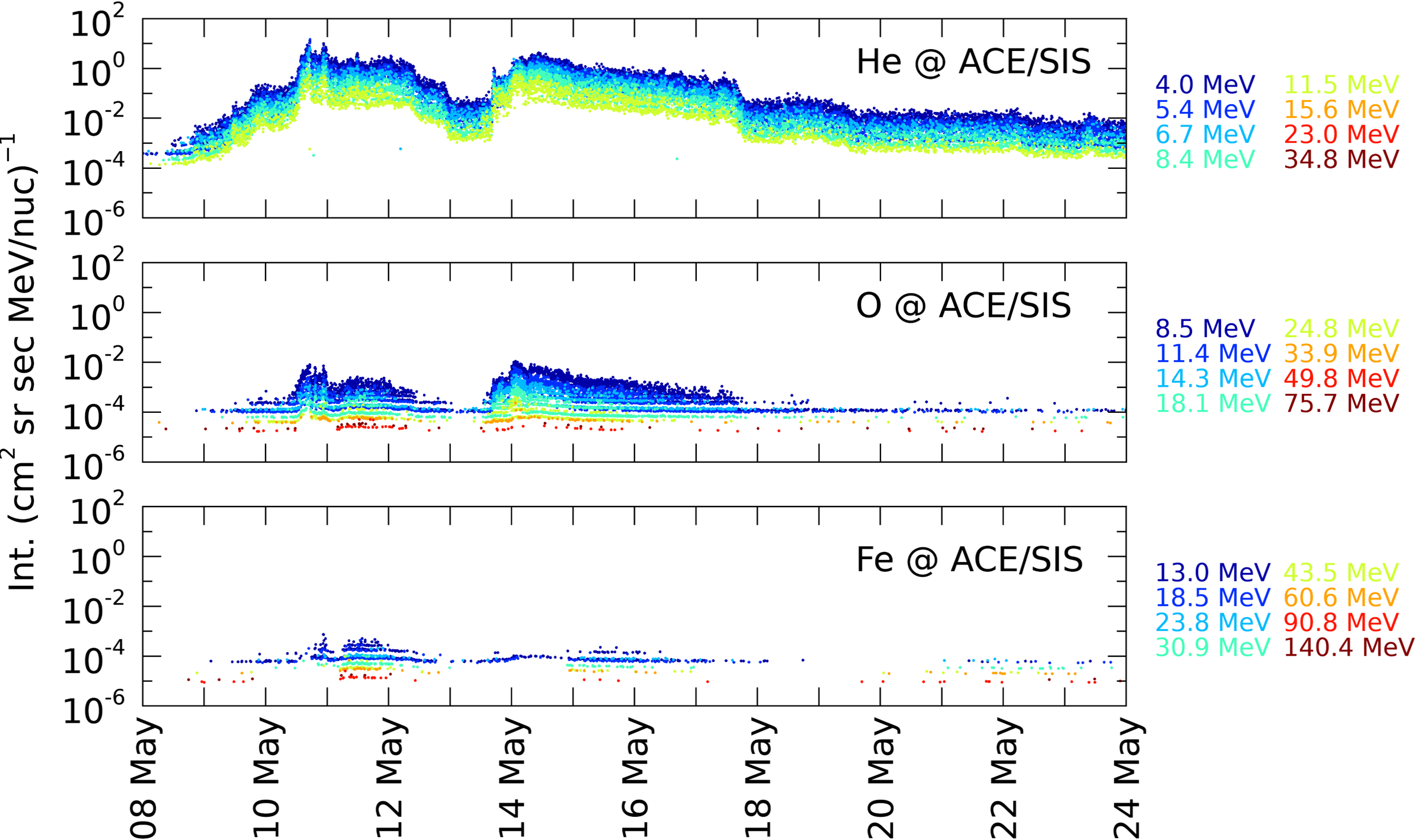}
    \caption{Intensity plots measured by ACE/SIS for: (1) helium, (2) oxygen, and (3) iron at 256 second cadence.}
    \label{fig:ace_sis}
\end{figure}

\begin{figure}[htbp]
    \centering
    \includegraphics[width=1.0\linewidth]{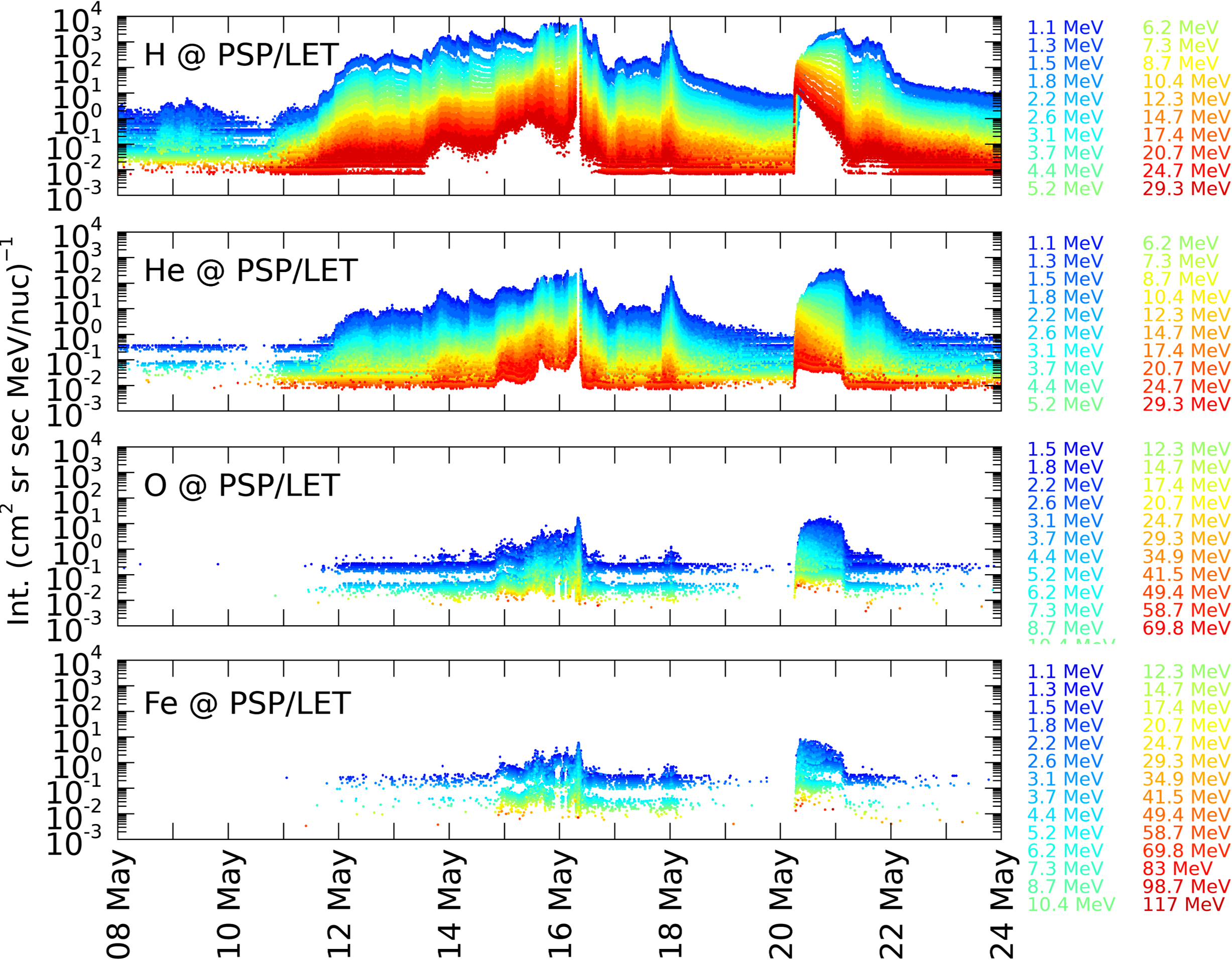}
    \caption{Intensity plots measured by PSP/LET A for: (1) protons, (2) helium, (3) oxygen, and (4) iron at 1 minute cadence. The cut out in the H and He intensities at lower energies on 16 May 2024 is due the instrument being in \change{dynamic threshold}{DT} modes \citep{cohen2021}.}
    \label{fig:psp_let}
\end{figure}

Intensity plots, shown in Figures \ref{fig:stereo_let}, \ref{fig:ace_sis} and \ref{fig:psp_let}, display the intensity at a given energy vs. time throughout the entire 8 to 24 May 2024 superstorm period. It is evident that there is large variation in \change{intensity as well as}{in both intensity and} composition throughout the entire superstorm period, \change{which differences for}{with distinct patterns observed at} each spacecraft.

\begin{figure}[htbp]
    \centering
    \includegraphics[width=0.99\linewidth]{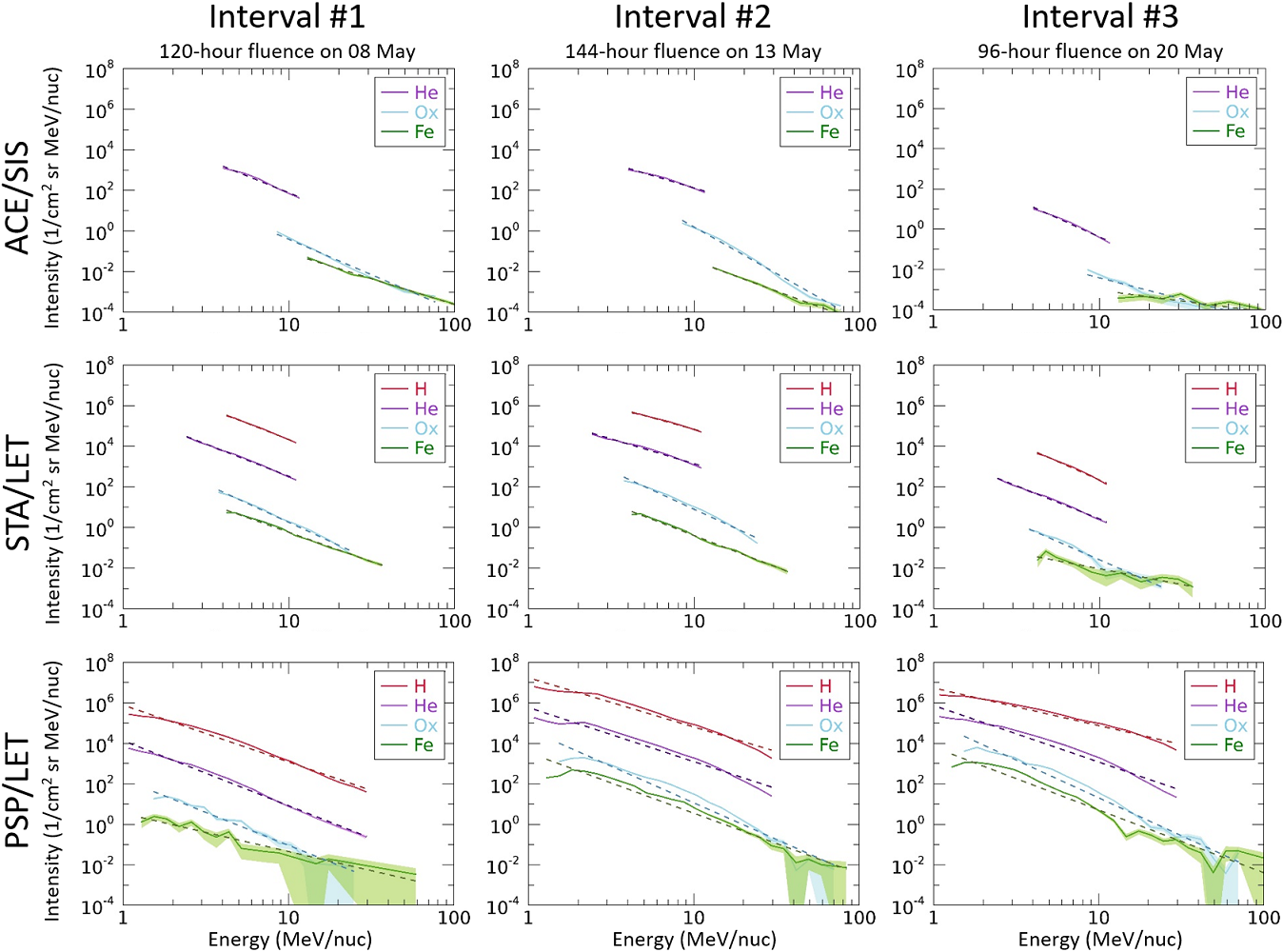}
    \caption{Fluence spectra from top to bottom rows: ACE/SIS, STA/LET, and PSP/LET for each of the intervals identified in Figure \ref{fig-storms}. The shaded area indicates uncertainty\add{ and the dashed lines are power-law fits with indices listed in Table {\ref{tab-indices}}}. (Left column) Interval 1 includes ion fluence for 8-13 May, (middle column) Interval 2 for 13-20 May, and (right column) Interval 3 for 20-24 May. An animation of 12 hour fluences, updating each hour is crucial for observing short duration changes not apparent in this stationary figure, it is located at {\url{https://youtu.be/lH7xk3TrLfk}}.}
    \label{fig-fluence}
\end{figure}

Figure \ref{fig-fluence} shows the ion fluence spectra for H, He, O, and Fe  calculated for STA/LET and PSP/LET by integrating the energy intensity over time for each of the 3 intervals. ACE/SIS fluence spectra are processed the same way but without protons. In the case of PSP/LET and PSP/HET, only the sunward-facing A side was used in order to match the same orientation as the instruments on ACE and STA.

\begin{table}
  \caption{\add{Indices for fit lines in Figure {\ref{fig-fluence}} fluence spectra, organized by ion species and interval, with the energy per nucleon (MeV/nuc) range used to calculate fits listed in parentheses.}}
  \centering
  \begin{tabular}{l l c c c}
    \hline
     & & Interval 1 & Interval 2 & Interval 3 \\
    \hline
    H & STA/LET  & -3.19 (4.2,11.0) & -2.26 (4.2,11.0) & -3.36 (4.2,11.0) \\
    H & PSP/LET  & -2.81 (1.1,29.3) & -2.43 (1.1,29.3) & -1.86 (1.1,29.3) \\
    \hline
    He & ACE/SIS & -3.29 (4.0,11.5) & -2.43 (4.0,11.5) & -3.71 (4.0,11.5) \\
    He & STA/LET & -3.17 (2.4,11.0) & -2.45 (2.4,11.0) & -3.25 (2.4,11.0) \\
    He & PSP/LET & -3.21 (1.1,29.3) & -2.67 (1.1,29.3) & -2.80 (1.1,29.3) \\
    \hline
    O & ACE/SIS  & -3.50 (8.5,75.7) & -4.61 (8.5,75.7) & -2.13 (8.5,75.7) \\
    O & STA/LET  & -3.75 (3.8,23.8) & -3.74 (3.8,23.8) & -3.56 (3.8,23.8) \\
    O & PSP/LET  & -3.26 (1.5,24.7) & -3.62 (1.5,69.8) & -3.73 (1.5,69.8) \\
    \hline
    Fe & ACE/SIS & -2.53 (13.0,140) & -3.00 (13.0,90.9) & -1.08 (13.0,140) \\
    Fe & STA/LET & -2.89 (4.2,36.3) & -3.14 (4.2,36.3) & -1.56 (4.2,36.3) \\
    Fe & PSP/LET & -1.89 (1.3,58.7) & -3.00 (1.3,83.0) & -3.10 (1.3,117) \\
    \hline
  \end{tabular}
  \label{tab-indices}
\end{table}

The fluence spectra shown in Figure \ref{fig-fluence} are for different durations due to the varying length of each interval marked in Figure \ref{fig-storms}. Thus, overall fluence is high for Interval 2 since it had sustained high intensities over a longer period of time, indicating that there were more SEPs during this period at all three spacecraft. An animation{\footnote{\url{https://youtu.be/lH7xk3TrLfk}}} of Figure {\ref{fig-fluence}} that covers a consistent 12 hour period that updates each hour is most suitable for observing the dynamic evolution of the fluence spectra occurring at each spacecraft.

The Interval 1 period shows the softest fluence spectra across all ions, with the highest fluence occurring at ACE and STA across the same energy ranges. Note that PSP/LET shows high uncertainty for Fe ions due to low statistics above \app10 MeV/nuc. The O and Fe fluence spectra for both ACE and STA are most similar in this period compared to the other intervals.

Interval 2 shows a slight hardening of the spectra, relative to Interval 1, most readily visible at ACE/SIS and STA/LET, but their relative fluence has decreased compared to that of PSP/LET. Additionally, O and Fe fluence spectra at ACE/SIS and STA/LET display a notably lower Fe/O ratio compared to Interval 1, while at PSP/LET the ratios were similar for both intervals.

Interval 3 has different relative fluence, with PSP/LET displaying a hard proton fluence spectra possibly due to the strong X16.5 flare and good connectivity to AR 13664. Additionally, the abundance of Fe and O are highest at PSP/LET relative to protons. The softening of the fluence spectra for ACE/SIS and STA/LET might be explained by the reduced connectivity with AR 13664 after it rotated to the far side of the Sun. 

For all intervals, there are no apparent spectral break energies over the measured energy ranges, thus we fit power laws to each spectra \change{resulting in the indices shown in Figure {\ref{fig-fluence}}}{in Figure {\ref{fig-fluence}} and have listed the resulting indices in Table {\ref{tab-indices}} for each ion species and interval}. The O and Fe spectra are described by significantly different power laws. Since Fe has a harder spectrum, it leads to an energy-dependent Fe/O ratio discussed below.

\section{Fe/O analysis}\label{fe_o_analysis}

\begin{figure}[htbp]
    \centering
    \includegraphics[width=0.9\linewidth]{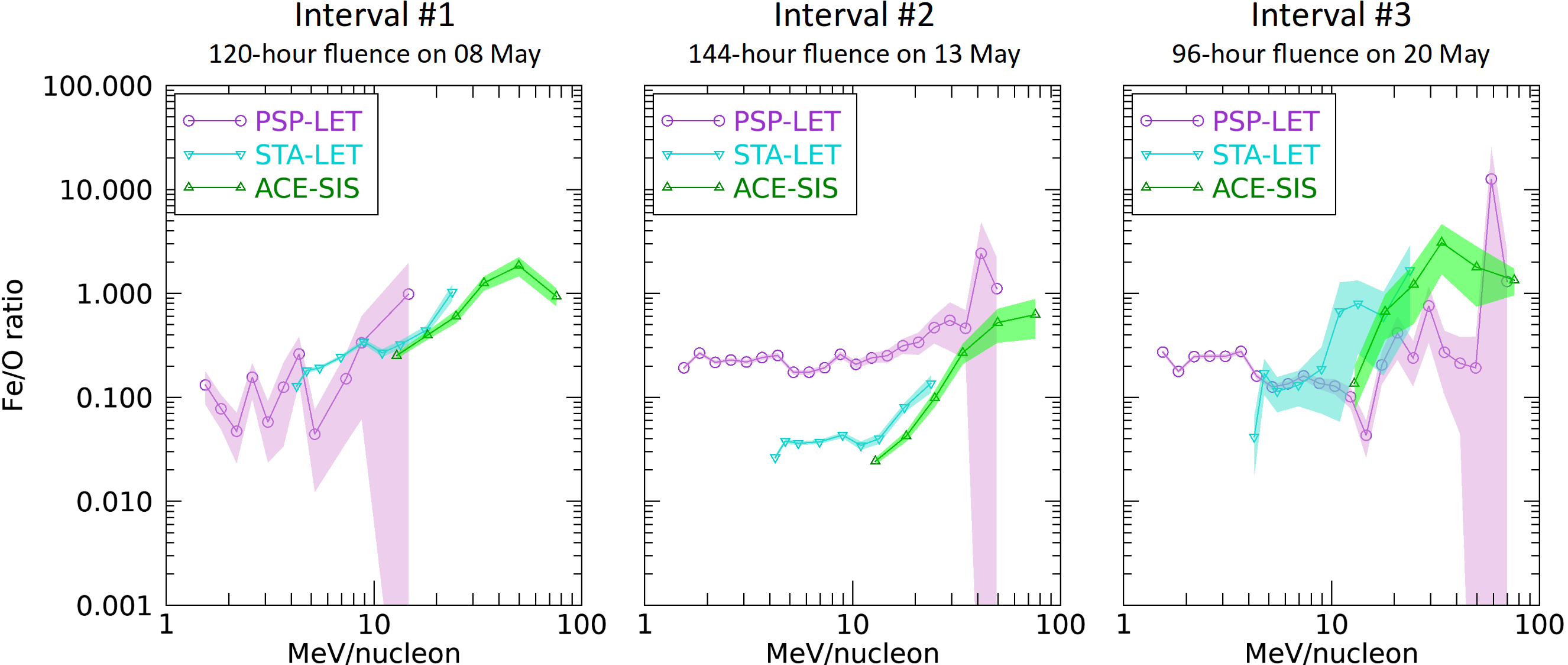}
    \caption{Fe/O fluence ratio starting at (left) 8 May for Interval 1, (middle) 13 May for Interval 2, and (right) 20 May for Interval 3. Within each plot, ACE/SIS is green, STA/LET is cyan and PSP/LET is violet. The shaded area indicates uncertainty. The ratio is taken over 120, 144, and 96 hours for each of the three intervals. To observe shorter, 12-hour fluence ratio changes, an animation located at {\url{https://youtu.be/XUctxb_NriQ}} updates each hour throughout the study period.}
    \label{fig-feox_ratio}
\end{figure}

The average Fe/O ratio for 5 to 12 MeV/nuc in SEP events is 0.134 as reported by \citet{reames1995} during solar cycle 21 and 22. Higher Fe/O ratios have been reported at lower energies as 0.236 in interplanetary shocks at 750 KeV/nuc by \citet{desai2003}, as 0.404 in "gradual" SEP events at 385 keV by \citet{desai2006}, and as 0.950 in "impulsive" SEP events at 385 keV by \citet{mason2004}. 

Figure \ref{fig-feox_ratio} shows the Fe/O ratio for each of the intervals as a function of energy for each spacecraft. Interval 1 clearly shows the highest uncertainties at PSP, as the event is much weaker with poor statistics, likely a result of poor magnetic connection to AR 13664 which is well eastward of the spacecraft on the Earth-facing side of the Sun. ACE and STA, each with low uncertainties, show similar Fe/O ratios with ACE values above 1.0 near 50 MeV/nuc. Both ACE and STA Fe/O abundances strongly increase with energy.

Interval 2 shows low uncertainties across all three spacecraft, except for PSP above 40 MeV/nuc, through nearly the entire energy range, with PSP having a relatively energy-independent Fe/O ratio of \app0.2 between 1.5 and \app20 MeV/nuc. Both STA and ACE have significantly lower Fe/O ratios than in Interval 1, with less consistency between the spacecraft from 12 to 25 MeV/nuc and ACE/SIS being the slightly lower of the two. From \app4 to 30 MeV/nuc, STA/LET and ACE/SIS are significantly lower than PSP/LET at the same energies. As in the previous interval, the general trend for ACE and STA is that Fe/O increases with energy, with the highest energies of ACE/SIS at an Fe/O ratio of 0.6 enhancement.

In Interval 3 the data are more uncertain for STA due to lower intensity measurements, although the general trend of Fe/O increasing with energy is still observed and surpasses Fe/O \gt0.5 above 20 MeV. The observed energy dependence of the Fe/O fluence spectrum is also roughly consistent between ACE/SIS and STA/LET, with the former reaching Fe/O \gt1 above 30 MeV. \add{In Figures {\ref{fig:ace_sis}} and {\ref{fig:stereo_let}}, the counts at STA and ACE are relatively low, which introduces larger uncertainties in the Fe/O determination reflected in Figure {\ref{fig-feox_ratio}}. Nevertheless, the measured ACE/SIS and STA/LET intensities exceed the pre-event background levels and align temporally with the 20 May event onset, supporting the interpretation that they are representative of the SEP event while recognizing the limited statistical confidence.} In contrast, PSP exhibits a slight decrease in Fe/O with energy below 10 MeV/nuc and has a strong dip in the Fe/O ratio at 12.3 MeV/nuc, visible in the fully integrated fluence of Figure \ref{fig-feox_ratio}c, which persists for just under 17 hours at the beginning rapid onset of the 20 May SEP event before recovering to values qualitatively similar to that of Interval 2.

\begin{figure}[htbp]
    \centering
    \includegraphics[width=0.9\linewidth]{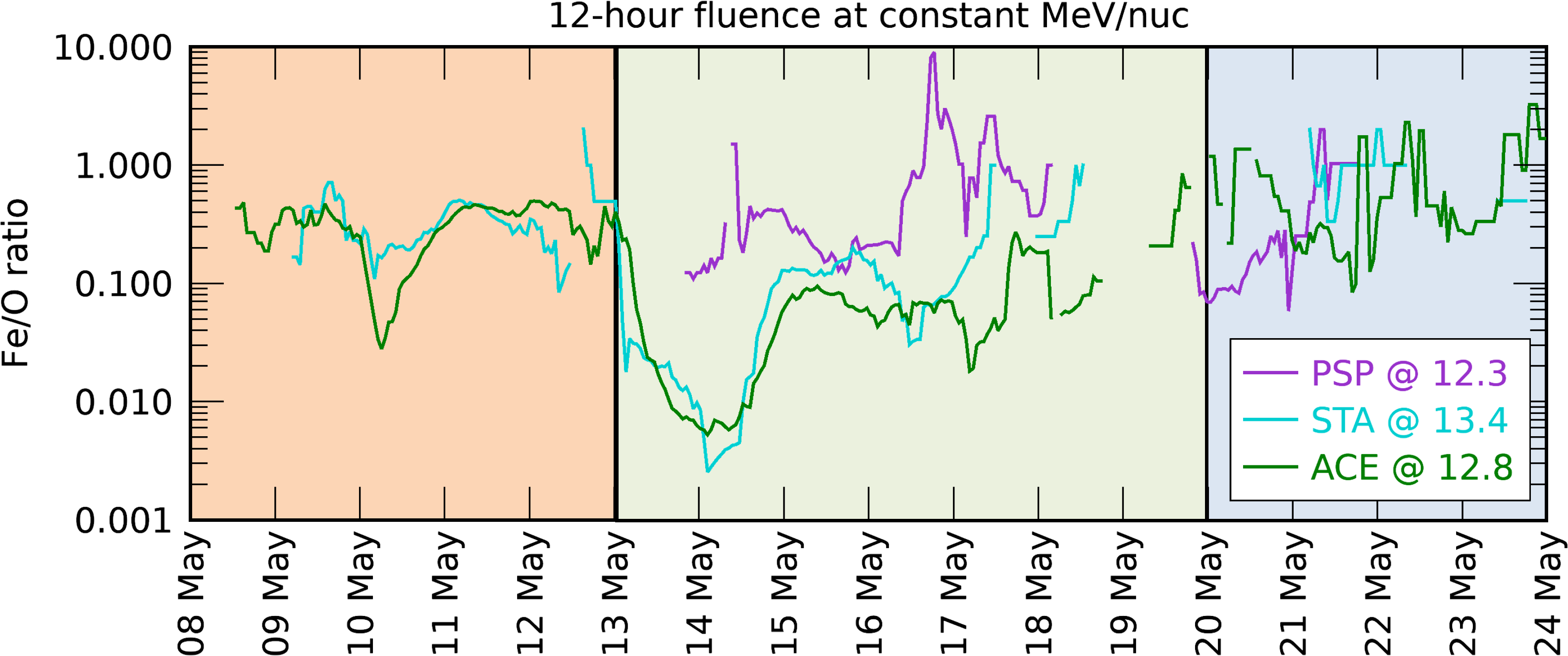}
    \caption{12 hour Fe/O ratios for similar energies at each spacecraft. ACE/SIS at 12.8 MeV/nuc, STA/LET at 13.4 MeV/nuc and PSP/LET at 12.3 MeV/nuc. The light orange, light green, and light blue time periods correspond to Intervals 1, 2, and 3, respectively. An animation is crucial for observing the short duration changes and uncertainty for each given energy. The animation combines Figures \ref{fig-feox_ratio} and \ref{fig-feox_hourly} and updates each hour and is located at {\url{https://youtu.be/XUctxb_NriQ}}.}
    \label{fig-feox_hourly}
\end{figure}

To study the temporal behavior of the Fe/O ratio, Figure \ref{fig-feox_hourly} plots Fe/O at similar energies between 12.3 to 13.4 MeV/nuc for all three spacecraft integrated over 12 hours, at sliding 1 hour intervals. This is best observed as an animation{\footnote{\url{https://youtu.be/XUctxb_NriQ}}} that updates each hour to view the context of these single energy measurements within the full energy range and also observe the uncertainties at a given time.

Initially, on 9 May, ACE/SIS and STA/LET have similar Fe/O ratios around \app0.4. A steep decrease occurs on 10 May when the merged CMEs arrive at Earth. While STA/LET Fe/O ratio only reaches 0.1, the ACE/SIS ratio undergoes a much lower dip to 0.03, before returning to values similar to STA on 12 May.

14 May marks the largest decline in Fe/O for ACE and STA, reaching 0.006 Fe/O at the same time that PSP begins to measure Fe/O at this energy. The start of measurements at PSP coincides with AR 13664 beginning to rotate to the PSP's subsolar point. The most significant increase in Fe/O occurs on 16 May, when the largest B$_{total}$ jump of May 2024 was measured (see Figure \ref{fig-storms}), and reaches a value of nearly 10.0.

The final notable Fe/O change occurs on 20 May, when a strong flare occurred in association with a fast-rising SEP event at PSP. While the 0.1 Fe/O is fairly typical for SEP events, due to the assumed good magnetic connectivity with the X16.5 flare one would expect a much higher Fe/O ratio often associated with flare accelerated SEP events. When examining Figure \ref{fig-feox_ratio} for Interval 3, the small dip in Fe fluence spectra at 12.3 MeV/nuc could be driving the displayed ratio slightly lower than expected.

\section{Discussion}\label{discussion}

\subsection{Time profiles and evolving magnetic connection}\label{time_profiles}

Comparison of the SEP flux enhancements shown in Figure \ref{fig-storms} during the mid-May 2024 events suggests that ACE/SIS and STA/LET, both located near Earth, were magnetically well-connected to AR 13664 during the first half of the study period. PSP/LET, positioned over 90\degree\ east in Stonyhurst longitude from Earth, did not register notable flux enhancements until 11 May. If the early shocks associated with AR 13664 during the first three days of the observation period were relatively narrow, on the order of \app100\degree\ in longitudinal width, then PSP simply fell outside the effective shock-connected region \citep{cane2000} due to it being far west of AR 13664 as shown in Figure \ref{fig-solarmach}. So, the gradual increase in SEP flux observed at PSP on 11 May could reflect late magnetic connection to a more extended shock front, with connectivity occurring far from the solar surface and deeper into the interplanetary medium \citep{strauss2017}. The rapid-fire CME activity during this period \citep{hayakawa2025} likely contributed to increased solar wind turbulence, complicating particle transport. Elevated levels of magnetic field fluctuations can favor cross-field diffusion \citep{giacalone1999,chollet2011,giacalone2012}, further reinforcing the observed gradual SEP rise at PSP shown in Figure \ref{fig-storms}. In this context, the slow SEP flux increase \change{can be interpreted as an implication of weak cross-field transport and improving connectivity to an initially inaccessible shock source as AR 13664 rotated towards PSP.}{is consistent with weak cross-field transport and improving connectivity as solar rotation shifted PSP’s nominal Parker-spiral footpoint onto corotating, particle-filled flux tubes connected to the shock.}



\begin{figure}[htbp]
    \centering
    \includegraphics[width=0.65\linewidth]{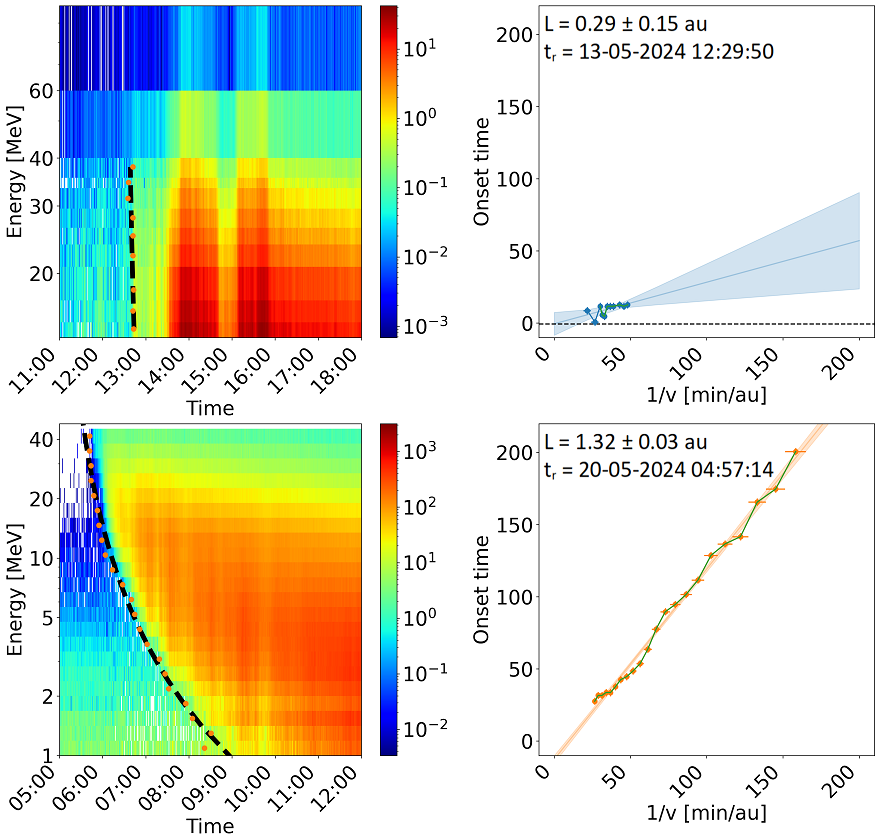}
    \caption{Proton intensity spectrograms for 7 hours at 1 minute intervals on (top row) 13 May by STA/HET and (bottom row) 20 May by PSP/LET combined with PSP/HET. The orange dots are automated fit points for the arrival of protons at a given energy. Velocity dispersion analysis plots of the spectrogram fit points, providing the path length (L) and particle solar release time ($t_{r}$).
    }
    \label{fig-vda}
\end{figure}

Velocity dispersion analysis (VDA) can be utilized as an additional diagnostic for assessing the evolving magnetic connectivity between a spacecraft's particle detector and AR 13664. By examining the onset times of SEPs across energy channels and assuming all particles were released from the same source at the same time, VDA is able to estimate both the initial SEPs release time and the apparent path length traveled \citep{xu2020}. However, Figure \ref{fig-vda}a shows the typical result from the highest energy particle detectors, STA/HET and PSP/HET, throughout most of the observation period in Intervals 1 and 2. Despite the qualitative appearance of rapid onset SEP events at STA on 13 May and other moments of abrupt intensity increases, when viewing long time series (e.g., Figure \ref{fig-storms}), the example VDA fit shown in Figures \ref{fig-vda}a suggests that nearly all intensity jumps in Interval 1 and 2 were poorly connected to their \remove{seed }particle source, as indicated by high uncertainty in its path length. The inability to fit VDA with reasonable uncertainty may imply substantial particle scattering during transport, potentially exacerbated by successive CMEs and turbulent solar wind conditions \citep{laitinen2010} during this mid-May 2024 period.


On 16 May, PSP/LET registered its peak proton intensity of the entire study period, an SEP enhancement that was intense enough to trigger \change{dynamic threshold}{DT} mode 2 \citep{cohen2021}, a protective mode designed to preserve live time and allow measurements during very high-rate periods of excessive particle flux. While the event appeared consistent with the arrival of a shock, VDA fitting was not possible even for this dramatic injection of SEPs, suggesting that SEPs were locally accelerated and not traveling from the same starting location. Additionally, none of the three aforementioned SEP event "pulses", on 10, 13, or 16 May, were temporally associated with a clearly identifiable flare from AR 13664 in a timeframe compatible with the arrival of high-energy protons measured at STA/HET or PSP/HET. This raises questions about the nature of SEP acceleration during this superstorm period, which does not seem to be driven directly by the high number of X-class flares \citep{stiefel2025}, but instead related to successive CMEs driving shocks that accelerate and inject energetic particles \citep{gosling1993}.


\begin{figure}[htbp]
    \centering
    \includegraphics[width=0.60\linewidth]{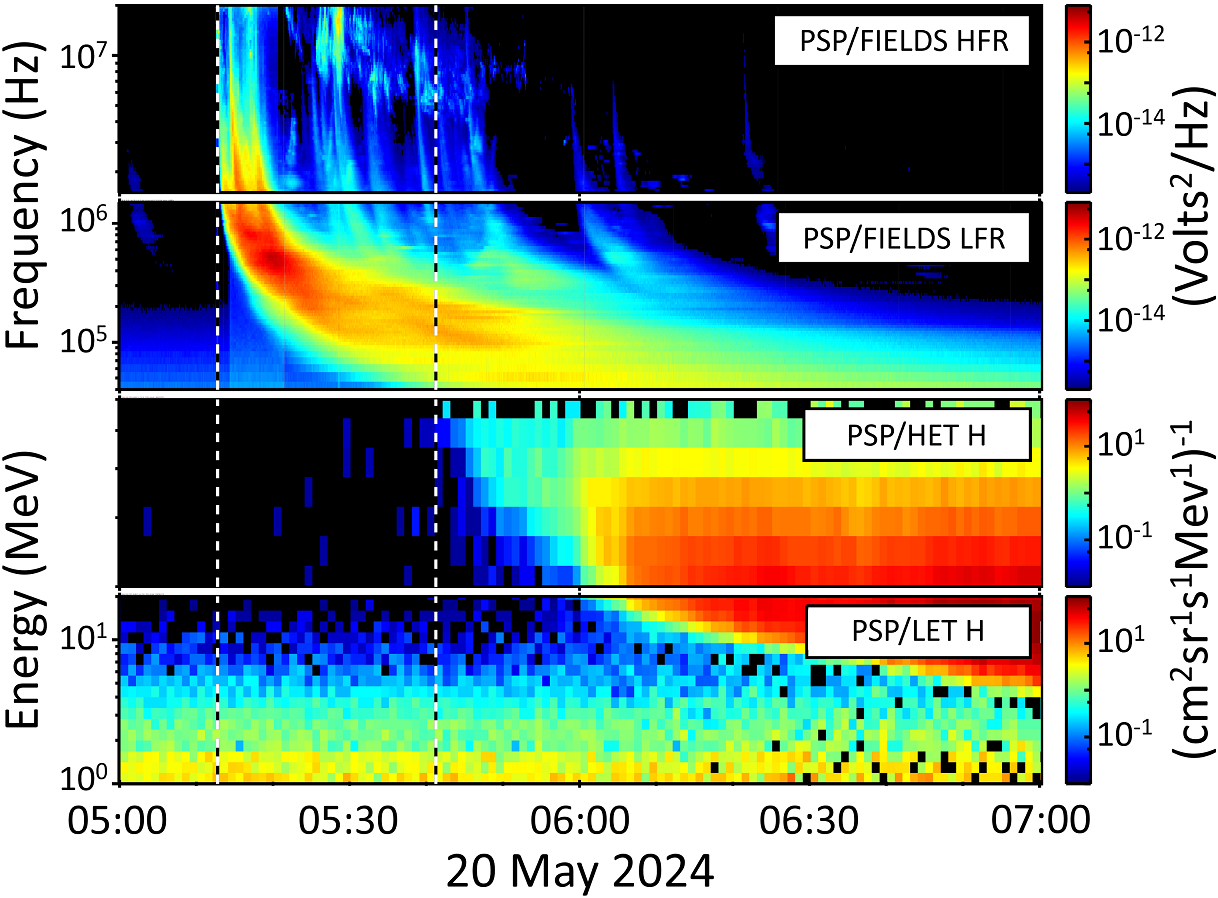}
    \caption{PSP time-series spectrogram data on 20 May 2024 of (Top 2 panels) radio emissions from FIELDS High and Low Frequency Receivers that show Type II/III radio bursts beginning at 05:12 and (Bottom 2 panels) proton intensity from \isois /EPI-Hi that shows the onset time of the SEP event at 05:41.}
    \label{fig-radio}
\end{figure}

In contrast, the SEP event on 20 May at PSP displays characteristics more indicative of a direct and well-connected flare-driven particle acceleration\add{, as exhibited by the enhanced Fe/O ratio at low MeV/nuc shown in the 12-hour fluence animation of Figure {\ref{fig-feox_hourly}}}. A powerful X16.5 flare \change{and Type III radio burst}{generated Type II and III radio bursts, shown in Figure {\ref{fig-radio}}, which} occurred just 29 minutes prior to the arrival of the highest energy SEPs at PSP, and the VDA result shown in Figure \ref{fig-vda}b demonstrates a well-constrained fit with minimal evidence of scattering at the event onset. The correlation between the flare timing, radio burst, and SEP arrival is strong, although the inferred SEP path length of 1.32 AU is longer than expected for direct magnetic connection at 0.74 AU radial distance. Unlike the earlier bursts of SEPs, this observation supports the view that by 20 May, AR 13664 had rotated into an optimal longitudinal position for particle injection into PSP-connected field lines. Curiously, however, despite being associated with the most powerful flare observed during the PSP mission to date, the 20 May SEP event did not produce especially elevated proton fluxes and did not activate any \change{dynamic threshold}{DT mode} protections. This discrepancy between strong flare magnitude and milder-than-expected SEP intensity at PSP's location is interesting, because Mars experienced the strongest ground level event in its entire rover era while only 13\degree\ longitude ahead and at a nearly twice PSP's radial distance from the Sun \citep{posner2025}. 

\subsection{Spectral shapes}\label{spectral_shapes}

The shape of the fluence spectrum for individual ion species, whether hard or soft, can also be influenced by the ion's Q/M, particularly through its effect on the particle gyroradius. According to the relation $R = MV/Q$, where $R$ is the rigidity, $M$ the mass, $V$ the particle velocity, and $Q$ the charge, one can rearrange the equation to show that the kinetic energy \add{per nucleon }of a particle scales with Q/M at a given magnetic field strength. Consequently, in acceleration models such as DSA, this dependency suggests that each ion species should exhibit a fluence spectral break energy, the point at which further acceleration becomes inefficient and the power-law spectrum fails to fit the data adequately \citep{desai2016}. \add{This inefficiency arises because, as the CME-driven shock propagates outward, it weakens and turbulence lessens which lowers the maximum energy it can accelerate particles to and the higher rigidity particles escape more easily due to lowered turbulence} \citep{tylka2006} \add{, resulting in a spectral break at that energy transition.} So, heavier ions with lower Q/M, such as Fe, are expected to exhibit a spectral break at lower energies than lighter ions with higher Q/M, such as O \citep{li2021}. However, this prediction is not borne out in the observed spectra from the events under study. As seen in Figure \ref{fig-fluence} and its associated animation, there is no clear evidence of a spectral break in the Fe fluence spectrum within the analyzed energy range. Conversely, the only potential spectral break appears at \app12 MeV/nuc for O ions during the Interval 3 STA/LET observation, although a lack of statistics at higher energies makes it difficult be certain. This discrepancy challenges the DSA interpretation that below any such spectral break energy the spectral slope should be independent of ion species and reflects additional complexities in the acceleration mechanisms or transport effects during these superstorm events. A plausible explanation is that the spectral break energies for these ion species lie outside the observed energy ranges of 1 to 100 MeV/nuc, limiting our ability to identify roll-off behavior, but the observed spectra make it clear that advanced modeling of time-dependent conditions is needed. While outside of the scope of this study, preliminary analysis of the \lt1 MeV energy range measured by the EPI-Lo instrument on PSP/\isois\ also did not display any clear break energies.


If Fe and O ions had similar Q/M values due to complete ionization, shock acceleration models would predict similar efficiencies in accelerating both ion species to high energies \citep{morlino2011}. However, this is not what our data indicate. As shown in Figure \ref{fig-fluence}, the fluence spectra of Fe and O \change{diverge measurably}{intersect in Interval 1 and 3}, with Fe exhibiting a harder spectrum\remove{ or elevated relative abundance at higher energies}. \change{It's worth noting that in Interval 3, the He/H+ ratio at PSP/LET measures a H+ and He diverge with increasing energy.}{It's worth noting that in Interval 3 at PSP/LET in Figure {\ref{fig-fluence}}, the He/H ratio measures a harder H spectrum and the He spectrum softens with increasing energy.} The He/H ratio decreasing with energy is opposite of the trend you would expect \change{if}{from} Q/M related processes \change{were driving the increasing energy dependence of the Fe/O ratio. It's not clear why this discrepancy is occurring, but could point to other other non-Q/M influences that could be investigated in a future study.}{, such as particles escaping from the shock.}

\subsection{Fe/O energy-dependent variation}\label{Fe/o_energy_dependent_variation}

Our analysis of the Fe/O ratio reveals a consistent and yet somewhat anomalous pattern throughout the mid-May SEP events. As shown in Figure \ref{fig-feox_ratio}, each of the three major intervals exhibit an increasing Fe/O ratio with rising energy, an unexpected trend when compared to canonical models of solar wind composition and DSA. The increasing Fe/O ratio in Figure \ref {fig-feox_ratio} is a natural consequence of Fe maintaining higher relative fluence at high energies while O fluence starts high then drops off more steeply. The common assumption for Fe/O abundance in a particle seed source population is that the ratio is independent of energy. And for DSA, the Fe/O composition is expected to be energy independent below the spectral break energy and decreasing beyond it \citep{cane2003,li2005b,tylkalee2006,desai2016}. The observed behavior thus suggests that an alternative mechanism or unique condition may be at play during these events.

We consider an explanation for this unexpected energy dependence, which is strongly supported by the fluence spectra shown in Figure \ref{fig-fluence} by Interval 1 and 2, where the Fe fluence spectrum is flatter than O at energies above \app10 MeV/nuc \citep{tylka2003,mewaldt2005}. For Interval 1 and 2, this challenges DSA theory, which predicts similar spectral shapes for all ion species \citep{li2005b,desai2016}. Although, PSP in Interval 3 seems to display a flatter Fe/O ratio. 

Given that traditional models such as DSA appear insufficient to fully explain the observed Fe/O energy dependence, alternative approaches must be considered. One such possibility is proposed by \citet{li2005b}, which suggests that each CME may possess a distinct ion composition. In this model, the trailing edge of an expanding CME loop can carry a unique population of particles whose ion abundances differ significantly from both the ambient solar wind and the upstream CME environment. Extending this idea further, a dynamic interaction between multiple CMEs, such as a fast CME overtaking and merging with a slower one, could lead to a spatially complex and energy-dependent Fe/O ratio profiles. Suppose the fast CME-A with a higher proportion of Fe which are accelerated to a higher average energy by its CME driven shock, while the slow CME-B is more rich in O ions which are accelerated to a lower average energy by its CME driven shock. Upon merging, this combined CME structure could exhibit an Fe/O ratio that would be relatively flat at lower energies as it is dominated by CME-B's O spectral shape and rise with increasing energy as the influence of CME-A's Fe spectral shape becomes more pronounced as shown in Figure \ref{fig-feox_ratio}. If such a mechanism were driving the observed spectral trends, one would expect the Fe/O slope to eventually plateau at sufficiently high energies where the contribution from CME-B becomes negligible and only the Fe-rich population from CME-A remains. While our data do not display a flattening of the Fe/O slope at the highest measured energies, the transition may be occurring above 100 MeV/nuc.





The increasing Fe/O ratio with energy observed in this study is inconsistent with the expectations of \remove{idealized }DSA theory, which typically predicts ion species-independent power-law spectral shapes below a break energy and decreasing Fe/O ratios above it. However, \citet{schwadron2015} presents a more nuanced treatment, modeling particle acceleration in low coronal compression regions and at quasi-perpendicular shocks with finite lifetimes. These evolving acceleration sites remain efficient only for limited durations and affect the maximum energy that particles reach, in contrast to DSA models that assume acceleration can persist indefinitely. Such transient structures produce broken power-law spectra and species-dependent behavior due to the rigidity scaling of the acceleration process. The model in \citet{schwadron2015} shows that the energy per nucleon achieved by an ion is proportional to a power ${\frac{2\chi}{\chi+1}}$ of Q/M, where $\chi$ \change{represents the rigidity dependence of scattering mean free paths}{represents the power law coefficient}. Fe ions, which span a wide range of charge states, may attain significantly higher energies than O ions if accelerated with a sufficiently high Q. This offers an explanation for the observed increasing Fe/O ratio with the observed energy dependence in Intervals 1 and 2, shown in Figure \ref{fig-feox_ratio}, particularly when acceleration occurs at quasi-perpendicular structures in the low corona.

The application of this non-idealized DSA theory to the mid-May 2024 period supports a scenario in which multiple transient acceleration sites, such as shock fronts or compression regions associated with rapidly evolving CME structures, successively inject and re-accelerate SEP populations. This interpretation is consistent with the observed variability in the Fe/O ratio across spacecraft shown in figure \ref{fig-feox_hourly} and absence of clearly defined spectral breaks in the Fe or O fluence spectra shown in Figure \ref{fig-fluence}. The variability itself may serve as a diagnostic signature of short duration acceleration at multiple, spatially distinct sites. Importantly, the rapid-fire CME eruptions during Intervals 1 and 2 suggest a mechanism for generating a multiplicity of acceleration regions \citep{schwadron2015}. Under these conditions, the energy-dependent Fe/O ratio does not arise from any single flare or shock, but from a combination of high-Q, low-M ion acceleration at different sites, each contributing a distinct ion composition. \add{However, the increasing Fe/O ratio for ACE and STA in Interval 3 cannot be similarly explained and its origin remains unclear.}

\subsection{Stepwise increases in intensity as merged CMEs}\label{stepwise_increases_in_intensity_as_merged_cmes}

In Figures \ref{fig-storms} and \ref{fig:stereo_let}, during the early phase of Interval 1 (8 to 11 May), STA/HET and STA/LET display a series of stepwise increases in proton intensity. Notably, these increases occur without a corresponding decay phase, except for a brief reduction in intensity on 10 May that is quickly followed by another upward step in flux. This stepwise pattern suggests a complex injection or transport mechanism rather than a single SEP release. A similar trend in Figures \ref{fig-storms} and \ref{fig:psp_let} is seen in PSP/HET and PSP/LET data, though occurring slightly later, from 11 to 16 May, as AR 13664 rotates closer to PSP’s longitudinal position, improving its magnetic connectivity. \change{Our interpretation of these stepwise increases is that they represent the sequential passage of CMEs, each carrying enhanced SEP populations. As these CMEs sweep past the spacecraft, they deliver increasingly enriched SEP populations.}{We interpret these steps as the cumulative effect of a tightly spaced sequence of CME-driven shocks where each successive passage injects or re-accelerates additional SEPs, producing stepwise increases in flux as they pass the spacecraft.}

Supporting this idea, \citet{hayakawa2025} presented observations of interplanetary scintillation (IPS) enhancements at 327 MHz near Earth, which are associated with dense ion parcels entrained within CME structures. These enhancements are thought to form when fast CMEs compress background solar wind, generating dense, turbulent regions that correlate with SEP-rich environments \citep{iwai2019}. The merger of successive CMEs has been shown to significantly amplify geomagnetic activity by producing compressed solar wind structures \citep{scolini2020} and stronger shocks \citep{koehn2022}. The extreme geomagnetic conditions observed in mid-May 2024, characterized by widespread aurora and elevated proton intensities over several days, suggest that such CME mergers were likely occurring. Thus, the in situ particle fluxes presented in \change{this study}{Intervals 1 and 2} should be considered as the integrated result of multiple CME interactions, rather than isolated, single-event SEP injections\change{.}{, such as in Interval 3.}

This \add{stepwise} interpretation lends further support to the \citet{li2005b} model, which proposes that CME structures can exhibit varying ion abundances. The enhancement in SEPs in Fe-rich populations at higher energies may arise from the merged CME due to re-accelerated SEPs with differing compositional signatures. This is consistent with both the stepwise increases in proton intensity observed in Figure \ref{fig-storms} and the unusual Fe/O ratio trends shown in Figure \ref{fig-feox_ratio}.

\subsection{Flare contributions and shock geometries}\label{flare contributions and shock geometries}

The increasing Fe/O ratio with energy can also be due to the contribution of flare-accelerated particles, either directly from or as part of a remnant suprathermal population subsequently reprocessed by CME-driven shocks. \add{An increase of Fe/O with energy is expected if CME shocks re-accelerate a mixed seed population that includes flare suprathermals with enhanced Fe and a harder spectrum, so their contribution rises at higher energies. Quasi-perpendicular shocks impose higher injection thresholds and Q/M-dependent cutoffs, favoring higher-rigidity Fe at a given energy per nucleon }\citep{jokipii1987,mason1999,tylka2006,desai2016}. \citet{cane2003} suggested two distinct components in large SEP events: an initial flare-associated population, often enriched in Fe ions and characterized by rapid onsets with elevated Fe/O ratios, followed by a shock associated population with more solar wind-like composition. \citet{tylka2006} proposed that the variability in high-energy Fe/O among large SEP events could be attributed to shock geometry and a mixed seed particle population. In particular, they suggested that quasi-perpendicular shocks preferentially accelerate higher energy flare suprathermals, allowing their unique compositional signatures, such as enhanced Fe/O and high Fe charge states, to be maintained after acceleration, while quasi-parallel shocks accelerate lower energy seed particles where the Fe/O ratio is more like that of the solar wind. 

Direct flare contribution to the SEP population during the mid-May 2024 period is one possible explanation for the observed Fe/O behavior. This mechanism is most applicable when the spacecraft is magnetically well-connected to the flaring active region. For instance, it could explain Interval 3 at PSP or possibly Intervals 1 and 2 at ACE or STA. However, it cannot account for the Fe/O behavior seen during other intervals where magnetic connectivity to AR 13664 was poor. Alternatively, the presence of a flare-related suprathermal seed population, subsequently re-accelerated by CME-driven shocks, could also explain the increasing Fe/O ratios with energy. This scenario is consistent with the energy-dependent fluence spectra shown in Figure \ref{fig-fluence} and the Fe/O ratio trends in Figure \ref{fig-feox_ratio}. While we have evidence suggesting a flare suprathermal component was present, we do not know the shock geometries during acceleration. Without knowledge of whether the shocks were quasi-perpendicular, the viability of this explanation is unknown. Detailed modeling of the shock structures and their evolution will be required to fully assess whether re-acceleration of flare-origin seed particles can account for the mid-May 2024 superstorm period.

\remove{Finally, it is worth acknowledging that transport effects may also contribute to the complex Fe/O patterns observed during the May 2024 SEP events. Processes such as rigidity-dependent scattering, drift effects, and magnetic field line meandering can all influence the differential propagation of ion species through the heliosphere (Mason et al., 2012; Dalla et al., 2013; Fraschetti, 2021; Laitinen et al., 2023). These effects can act independently of, or in combination with, the source mechanisms discussed above, leading to spectral distortions, delayed onsets, and compositional gradients that are not easily attributed to acceleration physics alone. While disentangling the influences of source vs. re-acceleration vs. transport remains a challenge, the observed variations in Fe/O across energy and time suggest that many processes are likely at play in shaping the SEP events recorded during this superstorm period.}

\subsection{\add{Effect of transport processes}}

\add{Finally, it is worth acknowledging that transport effects may be shaping the observed SEP intensities and complex Fe/O patterns during the May 2024 SEP events. Processes such as rigidity-dependent scattering, drift effects, and magnetic field line meandering can all influence the differential propagation of ion species through the heliosphere} \citep{mason2012,dalla2013,fraschetti2021,laitinen2023}. 

\add{Rigidity–dependent scattering, focusing, and adiabatic deceleration can imprint time and energy trends in ion ratios even when the source composition is fixed. In particular, comparing Fe and O at energies where their rigidities match largely removes the often observed Fe/O enhancement during the early portion of an event, indicating that much of the short–timescale variation reflects adiabatic deceleration rather than changing source composition} \citep{mason2012}. \change[rev2]{Adiabatic deceleration is the loss of SEP energy caused by the much slower solar wind, as SEPs work against the solar wind as they propagate outward and their energy per nucleon decreases with time and distance regardless of collisions.}{Adiabatic deceleration of SEPs is a well-known and important charged-particle transport process in the heliosphere }\citep{jokipii1971,kota2000}. Since Fe and O at the same energy per nucleon have different rigidities, and thus different scattering histories and focusing, this gradual energy loss can produce apparent early Fe/O differences that aren’t due to changing source composition. When you compare Fe and O at matched rigidity (rather than matched energy), much of that transport-induced high Fe/O ratio disappears. The transport effect described in \citet{mason2012} \add{suggests that we should see a drop in the Fe/O at the initial onset of an SEP event, which may be behind the decrease on 10 May in Interval 1 and 13-14 May of Interval 2 in Figure {\ref{fig-feox_hourly}}, with the latter shown far more dramatically in the 12-hour sliding time window animation. It is unclear why no similar drop in Fe/O is observed during Interval 3.}

\add{With typical SEP charge states (e.g., Fe 10-14; O 6-8), Fe ions have a smaller Q/M compared to O ions, }\change[rev2]{so the pattern described in }{and }\citet{zelina2017}\add[rev2]{ identified a pattern of Fe/O ratios tending to peak early in an event and decreasing quickly, reflecting transport controlled by Q/M rather than rapid changes at the source.}\remove[rev2]{ suggests that the temporal behavior of Fe/O should trend down over time.}\add{ ACE, STA, and PSP have different connectivity to the source, which naturally modulates how strongly the Q/M ordering appears at each spacecraft as later connection should yield a weaker decrease in Fe/O in the early phase of an event. On 10 May in Interval 1, Figure {\ref{fig-feox_hourly}} shows a more pronounced early Fe/O decrease at ACE/SIS than STA/LET, possibly reflecting this local transport effect due to preceding CMEs distorting connectivity and creating a longer path to ACE. In the context of}\grn{} \citet{zelina2017}\add{, our results in Figure {\ref{fig-feox_hourly}} support transport, rather than changing source ion composition, as the primary driver of short-timescale Fe/O ratio evolution at similar MeV/nuc.}

\add{However, Figure {\ref{fig-feox_ratio}} more clearly shows the longer-term general trend of this superstorm period: an increase of Fe/O with energy. Given the abundance of X-class flares from AR 13664 }\citep{jaswal2025}\add{, the presence of flare-generated suprathermal seed ions is likely. Considering that quasi-perpendicular shocks preferentially inject more energetic ions, which should bias the higher energy Fe/O upward }\citep{tylka2005}\add{, this is mechanism could lead to the stepwise proton increases noted in Section {\ref{stepwise_increases_in_intensity_as_merged_cmes}} due to rapid-fire CME eruptions.}

\add{Transport effects that move particles across field lines can also shape composition and timing when gradient drifts in a Parker spiral preferentially affect ions} \change[rev2]{that at both higher energies and lower Q/M, providing a route to energy-dependent Fe/O shown.}{at both higher energies and lower Q/M, possibly explaining the energy dependent Fe/O ratios apparent in Figure }\ref{fig-feox_ratio}. SEP drifts are a non-negligible source of cross-field transport, especially at \app1 AU where drift magnitudes peak for relevant energies at STA and ACE \citep{dalla2013}\add{. Whereas, perpendicular diffusion is most efficient early in SEP events when intensity gradients are steep, and produce wide (\app120\degree) onsets that spread particles longitudinally and smooth longitudinal structure }\citep{strauss2017}\add{. Together, these two processes may account for the energy-dependent Fe/O ratio discussed in Section {\ref{Fe/o_energy_dependent_variation}} persisting across the stepwise rises at all spacecraft discussed in Section {\ref{stepwise_increases_in_intensity_as_merged_cmes}.}}

\add{These transport effects can act independently of, or in combination with, the source mechanisms discussed in earlier sections, leading to spectral distortions, delayed onsets, and compositional gradients that are not easily attributed to acceleration physics alone. While disentangling the influences of source vs. re-acceleration vs. transport remains a challenge suited best to detailed modeling, the observed variations in Fe/O across energy and time suggest that many transport processes are likely at play in shaping the SEP events recorded during this superstorm period.}

\section{Conclusion}\label{conclusion}

The SEP events of mid-May 2024, driven by active region AR 13664, presented a unique opportunity to study longitudinally distributed ion composition during a historically geoeffective period at Earth. Using in-situ data from PSP/HET, PSP/LET, STA/HET, STA/LET, and ACE/SIS, we analyzed SEP fluxes and ion composition across a broad heliographic baseline in longitude. AR 13664's initial position near the sub-Earth point allowed both ACE and STA to measure intense SEP flux during the earliest storms, while PSP, separated by approximately \app90\degree\ in solar longitude, became increasingly connected as AR 13664 rotated westward.

Our observations revealed large Fe/O variability across spacecraft and energy ranges, with notable enhancements in Fe/O at higher energies inconsistent with traditional DSA predictions. Stepwise proton intensity increases at both STA and PSP suggest a sequence of CME passages. From 8 to 19 May, stepwise proton intensity increases and delayed SEP onsets supported a complex sequence of CME passages and evolving magnetic connectivity to solar sources. Whereas, the 20 May SEP event, associated with the strongest X-class flare of the PSP mission thus far, displayed good VDA fits with low evidence of scattering and a reasonable path length of 1.32 AU which exhibits how high flare magnitude does not necessarily translate to the most intense SEP event. Instead the most intense SEP fluxes occurred in conjunction with the passage of an exceptionally strong shock.

A series of interpretations were discussed with the stepwise increases in proton intensity being best explained by the sequential passage and merger of multiple CMEs, each carrying distinct SEP populations, which collectively enhanced intensities and contributed to the anomalous Fe/O ratio trends seen across the events. The increasing Fe/O ratio with energy during the May 2024 SEP events may have resulted from flare re-accelerated suprathermals by quasi-perpendicular CME-driven shocks or species-dependent transport effects which shape the ion compositions observed. Although, we need detailed modeling of the CMEs, particle acceleration, and transport to fully understand this complex period.

\section*{Open Research Section}
The spacecraft data used in this study are publically available and were obtained from these sources: The PSP/ISOIS data was obtained from \url{https://spp-isois.sr.unh.edu/}, the STA/LET and HET data were downloaded from the STEREO science center at \url{https://stereo-ssc.nascom.nasa.gov/}, and the ACE/SIS data are downloaded from the ACE Science center at \url{https://izw1.caltech.edu/ACE/ASC/}.

\section*{Conflict of Interest Statement}
The authors have no conflicts of interest to disclose.

\acknowledgments
We acknowledge the contributions of the Parker Solar Probe, STEREO, and ACE mission teams in collecting and providing data for this study. Parker Solar Probe was designed, built, and is now operated by the Johns Hopkins Applied Physics Laboratory as part of NASA's Living with a Star program, contract No. NNN06AA01C. Support from the LWS management and technical team has played a critical role in the success of the Parker Solar Probe mission. We thank the scientists and engineers whose technical contributions prelaunch have made the \isois\ instruments such a success. C.M.S. Cohen acknowledges additional partial funding from NASA grants 80NSSC22K0893, 80NSSC21K1327, 80NSSC20K1815, 80NSSC23K0543, 80NSSC24K0175, and 80NSSC21K1512. F. Fraschetti acknowledges partial funding from NASA under Grant 80NSSC21K1766 (Parker Solar Probe Guest Investigators). As always, we thank the Sun for providing the energy and particles necessary for this inquiry. 

%
%

\bibliography{refs.bib}

%
%
%
%
%

\end{sloppypar}

\end{document}